\documentclass[10pt,journal,compsoc]{IEEEtran}
\ifCLASSOPTIONcompsoc
  \usepackage[nocompress]{cite}
\else
  \usepackage{cite}
\fi
\usepackage{graphicx}

\ifCLASSINFOpdf
\else

\fi

\begin{document}
\title{Towards understanding the nature of direct functional connectivity in visual brain network}
% desired complexities of \author between the different types of papers makes a

\author{Debanjali Bhattacharya,
        and Neelam Sinha
        %~\IEEEmembership{Member,~IEEE,}
    
\IEEEcompsocitemizethanks{\IEEEcompsocthanksitem The authors are with International Institute of Information Technology (IIIT), Bangalore 560100, Karnataka, India.\protect\\
E-mail: debanjali.bhattacharya@iiitb.ac.in, neelam.sinha@iiitb.ac.in. }
%\IEEEcompsocthanksitem Neelam Sinha is with International Institute of Information Technology, Bangalore 560100, Karnataka, India.\protect\\
%E-mail: neelam.sinha@iiitb.ac.in}% <-this % stops an unwanted space
}

\IEEEtitleabstractindextext{%
\begin{abstract}

Recent advances in neuroimaging have enabled studies in functional connectivity (FC) of human brain, alongside investigation of the neuronal basis of cognition. One important FC study is the representation of vision in human brain. The release of publicly available dataset ”BOLD5000” has made it possible to study the brain dynamics during visual tasks in greater detail. In this paper, a comprehensive analysis of fMRI time series (TS) has been performed to explore different types of visual brain networks (VBN). The novelty of this work lies in (1) constructing VBN with consistently significant direct connectivity using both marginal and partial correlation, which is further analyzed using graph theoretic measures, (2) classification of VBNs as formed by image complexity-specific TS, using graphical features. In image complexity-specific VBN classification, XGBoost yields average accuracy in the range of 86.5\% to 91.5\% for positively correlated VBN, which is 2\% greater than that using negative correlation. This result not only reflects the distinguishing graphical characteristics of each image complexity-specific VBN, but also highlights the importance of studying both positively correlated and negatively correlated VBN to understand the how differently brain functions while viewing different complexities of real-world images.

\end{abstract}

}


\maketitle

\IEEEdisplaynontitleabstractindextext

% \fi
\IEEEpeerreviewmaketitle

\IEEEraisesectionheading{\section{Introduction}
\label{sec:introduction}}

\IEEEPARstart{N}{eurons} are the nerve cells which are considered as the most fundamental units of the brain and nervous system. There are approximately 86 billion neurons in human brain that are connected through approximately 150 trillion synapses which is responsible in transmitting electrical signals to other neurons \cite{ref1}. Thus, neurons form a neural circuit in brain that act as information pathways across different regions of the brain. Studies in understanding the complex characteristics of human brain have increased remarkably in past decades. In this regard, exploration of the brain neuronal connectivity patterns has been highlighted by neuroscientists since it reveals important information concerning the structural connectivity (pattern of anatomical links), functional connectivity (statistical dependency), and effective (causal interactions) connectivity of the human brain. In recent years, brain connectivity analysis has been the focus of many neuroscience researches \cite{ref2}.
During over last two decades several fMRI studies have been attempted in analyzing the functional brain connectivity pattern in patients with neurological disorders as well as in healthy individual with different cognitive load or during resting-state condition \cite{ref3, ref4,ref11,ref12,ref13}. However, limited studies have been performed in order to understand visual network connectivity in human brain. In this study, we focus on exploring functional architecture of brain during visual tasks using functional magnetic resonance imaging (fMRI) time series (TS) to investigate how vision represents in brain. \par
Brain FC network can be positive or negative depending on the associated pairwise correlation of BOLD TS. While the interpretation of positively correlated brain network is more instinctive, anti-correlation in brain network remains somewhat inscrutable. Thus, a growing attention has been paid in recent years to investigate the importance of anti-correlated brain network \cite{ref26, ref27,ref28,ref29,ref30}. In the present study, we have explored both positive and negative FC networks and their relationship to better understand the nature of neuronal interaction in brain sub-cortical regions, represents vision. 
The main contributions of this paper are:
\begin{itemize}
    \item Construction of visual brain network (VBN) is proposed by combining both marginal correlation and partial correlation estimates of fMRI TS to obtain consistently strong and direct (true) network connectivity for understanding brain FC properties underlying human vision.
\end{itemize}
\begin{itemize}
    \item A comprehensive graph-theoretical analysis is performed not only on brain networks having positive correlations but also on negatively-correlated brain networks to highlight the importance of considering anti-correlation in FC analysis.
\end{itemize}
\begin{itemize}
    \item Classification of image complexity-specific different VBNs using graph-based features, aiming to reveal the how differently human brain functions while looking into images of varying complexities.
\end{itemize}

\begin{figure}[t!]
\centering
\includegraphics[width=3.6in]{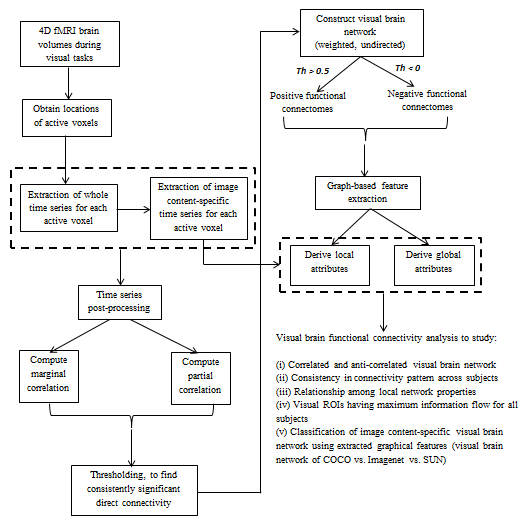}
%\DeclareGraphicsExtensions 
\caption{Block schematic of the proposed study.}
\label{fig_figure1}
\end{figure}

The block schematic of the proposed study is shown in Figure~\ref{fig_figure1}.

\section{Dataset description}
\label{sec:dataset}
\begin{figure}[t!]
\centering
\includegraphics[width=3.5in]{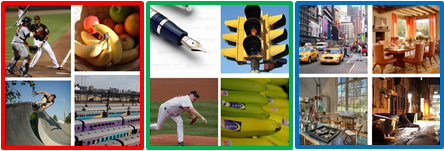}
%\DeclareGraphicsExtensions 
\caption{Sample images, taken from the three computer vision datasets having different complexities: \textit{Left:} COCO (\textit{Red box})- contains multiple objects and actions, \textit{Middle:} ImageNet (\textit{Green box})- contains single-focused object and \textit{Right:} SUN (\textit{Blue box})- contains indoor and outdoor scenes.}
\label{fig_imgs}
\end{figure}

The publicly available BOLD5000 dataset is used for this study \cite{ref14}. Four right-handed healthy volunteers (M:F 1:3)  were selected from Carnegie Mellon University with mean age 25.5 years. FMRI scans are acquired while viewing 5254 images of diverse complexities. These images for were selected from three classical computer vision database: \\
(i) \textit{Common objects in context (COCO)}, which is a standard benchmark dataset for multiple objects detection, segmentation tasks of complex indoor and outdoor scenes. 2000 images from COCO dataset are selected in the study. These multiple objects in COCO dataset depict interaction with other objects in realistic context (example, images depicting basic human social interactions).\\
(ii) \textit{Scene Understanding (SUN)} dataset that contains real-world scene images of indoor and outdoor environment and traditionally used for scene classification. The set of 1,000 scene images covering 250 categories were selected which inclined to be more panoramic, having no focus on specific object. \\
(iii) \textit{ImageNet} dataset, which is used for classification and localization of singular objects which are centered in real-world scenes. 1916 images were selected from ImageNet database that mainly focus on a single object in the picture. Unlike COCO and SUN images, the single object in the ImageNet images is mostly placed at center in order to distinguish it clearly from image background. The sample images from COCO, ImageNet and SUN are shown in Figure~\ref{fig_imgs}.  \par

Each functional session consisted of 9 to 10 runs where in each run 37 stimuli (images) were presented randomly to the participants. Each of the functional scanning sessions was roughly 1.5 hours long for all subjects. Further details regarding subject demographics with experimental paradigm, stimuli presentation, fMRI acquisition and data pre-processing procedures can be found in \cite{ref14}.

\subsection{Time series extraction from fMRI image volume}
\label{TS extraction}

In this paper, fMRI TS, which is the representation of BOLD intensity distribution over time, is utilized to study VBN. In this regard, two sets of TS are extracted (shown in Figure~\ref{fig_figure2}):
\textbf{SET-1:} the active voxel-specific whole TS as obtained from each run of a session and \textbf{SET-2:} the image complexity-specific TS from each of the three datbase: ImageNet, COCO and SUN- which is extracted from the whole TS. While the whole TS is used to examine the connectivity pattern of the general VBN, irrespective of what types of images were viewed by the participants, image complexity-specific TS is used to study the differences in VBNs, constructed while viewing images with different spatial complexities, as captured by three different dataset: ImageNet, COCO and SUN.

\subsubsection{Whole time series}
\label{sec:Whole time series}
In order to extract whole fMRI TS, first, it is necessary to have the information about active voxel locations in brain. In our study, SPM \cite{ref15} toolbox was used to get voxel locations from 4D \textit{(x,y,z and t)} fMRI data which were found to be significantly active during each run of the experiment. The active voxel locations are shown in Figure~\ref{fig_figure2}. Majority of these active voxels are found within 5 visual ROIs as defined in \cite{ref14} which are (i) the parahippocampal place area (PPA), (ii) the retrosplenial complex (RSC), (iii) the occipital place area (OPA), (iv) Early visual area (EV) and (v) lateral occipital complex (LOC). Some voxels that are found to be activated in other sub-cortical regions which are labelled as "others" in our study. From each run, the whole TS of length 37 (since, the no. of stimuli = 37, at each run) is extracted for each of the active voxel (Figure~\ref{fig_figure2}). The obtained voxel-specific TS whole is detrened and Z-score normalized before it is used for further analysis. 

\subsubsection{Image complexity-specific time series}
\label{sec:Image complexity-specific time series}
\begin{figure*}[t!]
\centering
\includegraphics[width=5in]{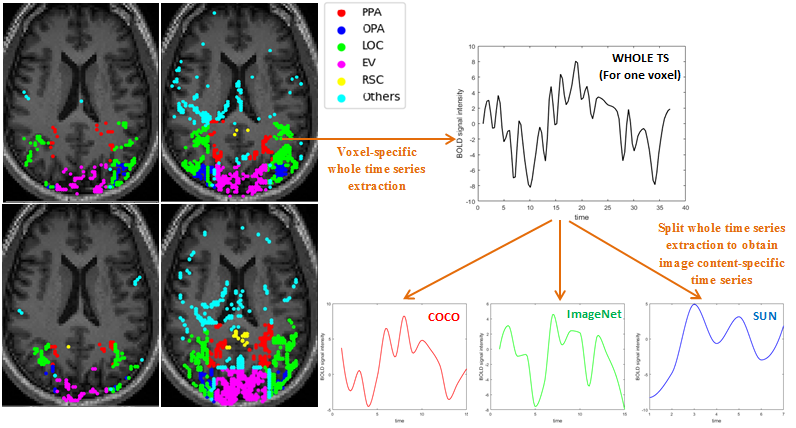}
%\DeclareGraphicsExtensions 
\caption{The MRI images in the figure (\textit{left}) illustrates voxel activation within brain while viewing images, for all four subjects. \textit{Right:} FMRI time series which is extracted from one active voxel is shown in \textit{Black} for one representative subject. Image complexity-specific time series is obtained by splitting the whole time series according to the BOLD intensity values of distinct image complexities (COCO, ImageNet and SUN), as seen by the participants at each time stamp. The three time series representing COCO, ImageNet and SUN as obtained from the whole time series are shown in \textit{Red}, \textit{Green} and \textit{Blue} color respectively.}
\label{fig_figure2}
\end{figure*}

In order to obtain the image complexity-specific fMRI TS, the whole TS is split into three parts according to the BOLD intensity values of images that represents a particular dataset (i.e whether it belongs to ImageNet or COCO or SUN). This is illustrated in Figure~\ref{fig_figure2}. Each of these three TS illustrates the BOLD intensity distribution while viewing images from COCO (Red), ImageNet (Green) and SUN (Blue). It is to be noted that, the length of these three TS that represents image complexities information of the three datasets are not same due to randomness in total number of image presentation to the participants from these three datasets.

\section{Functional connectivity analysis}
Cortical and sub-cortical regions in brain communicate with each other to process the information and thereby forms a distributed network. Identification and quantification of such inter-regional relationships within brain network can be captured in brain FC analysis \cite{ref1,ref2}. FC provides the information about the statistical dependencies between BOLD TS of two distinct brain regions. In the current study, MC and PC estimates of BOLD TS are utilized as FC measures to determine the connectivity strength between selected brain regions (voxel ROI).

\subsection{Estimating Marginal and Partial Correlation}
\label{sec: correlation}

Correlation analysis is the classical technique that has been widely used to study the nature of FC between two brain regions \cite{ref1,ref16,ref17,ref22,ref26,ref27,ref28,ref29,ref30}. It is seen that most of the studies used Pearson's correlation, that only captures the marginal association between network nodes (Marginal correlation or MC). However, brain FC analysis using only Pearson's correlation is not sufficient since it fails to capture the direct or true FC, if exists between network nodes. For example, significant correlations between two nodes, say, X and Y may occur due to their common connections to a third node, say Z, even if X and Y are not directly connected \cite{ref16,ref17}. Thus using MC, it is hard to differentiate the network edges that reflects true FC from the edges that reflects the connectivity caused by confounders. The statistical technique that has shown great potential in addressing this major issue is Partial correlation (PC) \cite{ref16,ref17,ref18}. 

\subsubsection{Partial Correlation}
\label{parcorr}
PC estimates correlations after regressing out spurious effects from all the other nodes in the network; making it a true measure of network connectivity. Thus, zero value in PC suggests an absence of direct connectivity between  network node pairs. Strong evidence from literature suggests that PC is one of the top techniques that outperform the traditional techniques and showed high sensitivity to find true functional connectivity between network nodes \cite{ref16,ref17,ref19}. 
PC of two TS between region 'i' and region 'j' at t is defined as the correlation between $X_{i}$ and $X_{j}$ conditioning on all the other nodes, i.e.:\\
\[ \rho_{ij} = corr(X_{i}X_{j}\mid X_{k}), where X_{k}: 1\leq k\leq M; k \neq i \neq j \]

The interpretation of the values of $\rho_{ij}$ not equal to zero is same as that with MC. However, when $\rho_{ij}=0$, node-i and j becomes conditionally independent given the other nodes. The PC of TS at any two nodes when number of nodes N$\geq$3 is defined as "the correlation of residuals obtained from linear regression of TS at first node with all other confounders and linear regression of TS at second node with all other confounders" \cite{ref16}. 
However, computation of PC using the regression approach is not appropriate in terms of computational time. Also it often fails if multicollinearity exists among TS. In such cases, instead of using linear regression, PC is estimated effectively from the inverse covariance matrix, also referred as the precision matrix. \par
%Let, $\Sigma$ is the $N \times N$ covariance matrix of two TS, and let, $\Omega_{N \times N}$ = $\Sigma^{-1}$ = $\omega_{ij}$ be the precision matrix, then, the PC between TS of node-i and node-j can be derived from precision matrix as,
%\begin{equation}
 %   \rho_{ij}=\frac{-\omega_{ij}}{\omega_{ii}\omega_{jj}}
%\end{equation}

Clearly, estimating PC using a precision matrix requires covariance matrix to be invertible. In neuroimaging applications, this is challenging when the number of active voxels exceed the number of observations (less number of fMRI scans). In such cases, derivation of the precision matrix needs more computational load and it may not be stable. There are many algorithms proposed in literature to overcome this difficulty \cite{ref16}. One such solution is to apply Moore-Penrose pseudo-inverse of the covariance matrix to directly estimate the precision matrix \cite{ref42}. In our study, Moore-Penrose pseudo inverse is used to estimate PC.

\section{Construction of VBN and Connectivity Matrix}
\label{sec:Construction of VBN and Connectivity Matrix}
\begin{figure}[t!]
\centering
\includegraphics[width=3.3in]{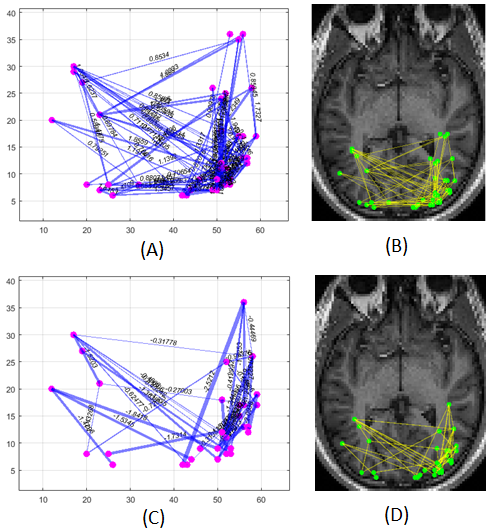}
%\DeclareGraphicsExtensions 
\caption{Construction of VBN from whole TS for one representative subject is shown. Figure (A) and (C) show graphical representation of networks with positive and negative edge strengths respectively. The respective mapping of network graph on fMRI image is shown in Figure (B) and (D).}
\label{fig_figure5}
\end{figure}

A network is a collection of nodes (vertices) and links (edges) between pairs of nodes (Figure~\ref{fig_figure5}).  
In context of brain network, nodes usually represent distinct brain regions, while the edges represent anatomical, functional, or effective connections between nodes\cite{ref2,ref22}. Thus, brain network connectivity can be mathematically represented as an undirected graph \textit{G(n,l)}, where, \textit{n} is a set of nodes and \textit{l} is a set of links, connecting node-\textit{i} and node-\textit{j}. Each link $l_{ij}$ in \textit{l} is associated with a weight that can be either positive or negative. In FC analysis of fMRI TS, the edge weights represent BOLD temporal correlations in brain activity that occur between two regions. Each network can be represented by network connectivity matrix or FC matrix (commonly known as Adjacency matrix) which is a symmetric matrix of size N $\times$ N, where, N is the number of nodes. Each $(i,j)^{th}$ entry in the FC matrix denotes the strength of edge connected between nodes \textit{i} and \textit{j}. In our study, the edge strengths are estimated using PC measure, used for constructing the VBN. \par
Since, correlation values vary from negative (-1) to positive (+1), the network constructed using correlation as FC measures can be categorized into two types: \textit{(i) Positive FC network (denoted by $FC^{+}$)} having positive edge strength and (ii) \textit{Negative FC network (denoted by $FC^{-}$)} having negative edge strength. In the context of brain FC analysis, positive correlation indicates that when activity in one region increases, increased activity is seen in the other region also. However, negative correlation indicates that when one region is more active, the activity in the other region decreases. In our study we have analyzed brain $FC^{+}$ and $FC^{-}$ networks using both the whole TS and image complexity-specific TS (as described in Section~\ref{sec:Whole time series} and Section~\ref{sec:Image complexity-specific time series}) in order to understand how vision is represented in human brain. The visualization of $FC^{+}$ and $FC^{-}$ brain networks are shown in Figure ~\ref{fig_figure5}.

\subsection{Properties of network graph}

An individual network measure obtained from network elements (nodes and edges) quantifies the connectivity profile which in turn can characterize several features of global and local brain architecture. In this section, we describe graph-based features that reveals functional integration and segregation aspects of brain network, aiming to understand the role of individual brain regions. These features are broadly categorized into two types: global features and local features.

\subsubsection{Graph-based global features}
\label{sec:globalfeatures}

Global measures of a network graph reveal functional segregation (degree to which network elements form specific modules or clusters) and functional integration (ability to combine segregated information from different modules of the network) of information flow within the brain network \cite{ref1,ref22}. Characteristic path length is one of the widely used measures that quantifies the properties of functional integration \cite{ref1,ref22}. On the other side, Modularity and Clustering coefficients are two most commonly used measures that quantifies global information segregation in brain networks \cite{ref22}. Four different global features which are extracted in the current study. These are defined as follows\cite{ref22}:\par

\textit{(1) Transitivity:} It is a  measure of the tendency of the network nodes to cluster together. High transitivity indicates that the network contains communities or groups of nodes that are densely connected internally. Transitivity is measured as,
\begin{equation}
    T=\frac{\sum_{i\in N}2t_{i}}{\sum_{i\in N}d_{i}(d_{i}-1)}
\end{equation}
where, $d_{i}$ is the degree of node 'i', i.e. $d_{i}=\sum_{j\in N}a_{ij}$, and \\
$t_{i}$ is the number of triangles around node 'i' which is computed as, $t_{i}=\frac{1}{2}\sum_{j,h\in N}a_{ij}a_{jh}a_{ih}$. Here, 'N' is the total number of nodes in the network

\textit{(2) Modularity:} It measures the strength of division of a network into modules or communities. Thus, networks with high modularity have dense connections between the nodes within modules but sparse connections between nodes in different modules. Modularity is measured as,
\begin{equation}
    M=\frac{1}{L}\sum_{i,j \in N}(a_{ij}-\frac{d_{i}d_{j}}{L})\alpha_{m_{i},m_{j}}
\end{equation}
where, $m_{i}, m_{j}$ are the module containing node 'i' and 'j' respectively,\\
$\alpha_{m_{i},m_{j}} = 1$, when $m_{i}=m_{j}$, otherwise it is equal to zero.
The total number of links is denoted by $L=\sum_{i,j \in N}a_{ij}$.\\
During computation of modularity index, each undirected link is counted twice to avoid ambiguity (i.e. $a_{ij}$ and $a_{ji}$ both are counted.

\textit{(3) Characteristic path length:} It is defined as the average shortest path length between all pairs of nodes in the network. This measure quantifies functional integration. Characteristic path length is computed as,
\begin{equation}
    \lambda=\frac{1}{N}\sum_{i\in N}\lambda_{i}=\frac{1}{N}\sum_{i\in N}\frac{\sum_{j\in N, j\neq i}\l_{ij}}{N-1}
\end{equation}
where, $\lambda_{i}$ is the average distance between node 'i' and all other nodes in the network. Here, $l_{ij}$ denotes the shortest path length of the network.  Ideally, $l_{ij}=\infty$ for all disconnected pairs i,j. In brain connectivity analysis toolbox, the default distance for all disconnected node pairs is assumed to be 1 (highest value) and the default diagonal distance of same node pair is assumed to be 0 (lowest value).

\textit{(4) Density:} Density defines the mean degree of the network. It is the ratio of actual connections ($AC=\sum_{i,j\in N}a_{ij}$) and the total number of all possible connections ($PC=\frac{N(N-1)}{2}$) in the network. Thus, density is computed as,
\begin{equation}
    D=\frac{2\sum_{i,j\in N}a_{ij}}{N(N-1)}
\end{equation}
Here, $a_{ij}=1$ when edges between node 'i' and 'j' exist, otherwise it is equal to zero.

\subsubsection{Graph-based local features}
\label{sec: localfeatures}

Local features are aimed to reveal the characteristics of each node in a network. In this work, nine such features are used to illustrate the local properties of VBN \cite{ref22}. The description of local features considered in this study are given below:\par
\textit{(1) Clustering coefficient:}
Clustering coefficient (CC) is the fraction of node’s neighbors that are neighbors of each other. Thus, it quantifies the abundance of connected triangles (clustered connectivity around individual nodes) in a network. CC is defined as,
\begin{equation}
    C=\frac{1}{N}\sum_{i \in N}C_{i}=\frac{1}{N}\sum_{i \in N}\frac{2t_{i}}{d_{i}(d_{i}-1)}
\end{equation}
where, $c_{i}$ is the CC of node 'i', $c_{i}=0$ for $d_{i}<2$. Here, $C$ denotes the mean CC.

\textit{(2) Degree:} Degree is the number of edges incident on a particular node. Degree of a node 'i' is computed as,
\begin{equation}
    d_{i}=\sum_{j \in N}a_{ij}
\end{equation}

\textit{(3) Betweenness centarlity:} The Betweenness centrality (BC) for each node is the number of shortest paths that pass through that specific node. A node with higher BC would have more control over the network as more information will pass through that node. It is computed as,
\begin{equation}
    b_{i}=\frac{1}{(N-1)(N-2)}\sum_{p,q \in N; p \neq q\neq i}\frac{l_{pq}(i)}{l_{pq}}
\end{equation}
$l_{pq}$ is the total number of shortest paths between node 'p' and node 'q', $l_{pq}(i)$ is the number of shortest path that pass through node 'i'.

\textit{(4) Local efficiency:} The local efficiency is a measure of the average efficiency of information transfer within neighborhoods and it is defined as the inverse of the shortest average path length of all neighbors of a given node among themselves. Local efficiency is computed as,
\begin{equation}
    \eta=\frac{1}{N}\sum_{i \in N}\eta_{i}=\frac{1}{N}\sum_{i \in N}\frac{\sum_{j,k \in N; j\neq i}a_{ij}a_{jk}[l_{jk(N_{i})}]^{-1}}{d_{i}(d_{i}-1)}
\end{equation}
where, $l_{jk(N_{i})}$ is the shortest path length between node 'j' and node 'k' that contains only neighbors of node 'i'.

\textit{(5) Eigen vector centrality:} Let, $A$ be the adjacency matrix of a network graph and $\lambda_{max}$ be the largest Eigen value of $A$ and $\vec{X}$ be the corresponding Eigen vector then, the $i^{th}$ component of $\vec{X}$ gives the Eigen vector centrality (EVC) score of the  $i^{th}$ node of the network. Thus, if $A\vec{X} = \lambda\vec{X}$, then, the EVC is computed as,
\begin{equation}
    e_{i} = x_{i} = \frac{1}{\lambda}\sum_{j=1}^{N}A_{ij}x{j}
\end{equation}
For any node 'i', the EVC score is proportional to the sum of scores of all nodes which are connected to 'i'. Hence, a network node will have high EVC if it is connected to many nodes that themselves have high EVC.

\textit{(6) Participation coefficient:} The distribution of connections of all edges of a node among separate modules is represented by participation coefficient (PCf). If the links of a specific node are entirely restricted to its own community (module), its PCf is 0. Nodes with a high participation coefficient (for example, connector hubs) are likely to facilitate global inter-modular integration. PCf is computed as,
\begin{equation}
    p_{i} = 1-\sum_{m \in M}\frac{d_{i}(m)}{d_{i}}^{2}
\end{equation}
where, M is the set of modules in the network, $d_{i}(m)$ is the number of links between node 'i' and all nodes in module 'm'.

\textit{(7) Diversity coefficient:} Diversity coefficient (DC) is an alternative measures of PCf. It measures a how well a node is connected outside to its own community. Nodes that have many connections to other communities will have higher diversity coefficient values.
\begin{equation}
    \delta_{i} = \sum_{m \in M}d_{i}(m)
\end{equation}

\textit{(8) Gateway coefficient:} This is similar to DC except that along with measuring diversity of the inter-modular connectivity of a specific node, gateway coefficient (GC) also measures how critical these connections are to inter-modular connectivity. Nodes which are solely responsible for inter-community connectivity will have higher GC values. In this measure, each term of the PCf is given a weight reflecting the importance of the connection. In our study, the weights are computed using the strength of the node. Thus, GC is computed as,
\begin{equation}
    g_{i} = p_{i}*s_{i}
\end{equation}
where, $p_{i}$ and $s_{i}$ denote the PCf and strength of node 'i' in the network.

\textit{(9) Strength:} It is defined as the sum of all neighboring edge weights which are connected to a particular node. So one can think strength as the weighted variants of the nodal degree. Strength of node 'i' is computed as,
\begin{equation}
    s_{i}=\sum_{i,j\in N, i\neq j}W_{ij}
\end{equation}
where, $W_{ij}$ is the edge weights of the adjacency matrix $a_{ij}$. 

\section{Results}

The BOLD5000 dataset as described in Section~\ref{sec:dataset} is used in this work in order to study the properties of different VBNs. To the best knowledge of authors, no work has been reported that utilized this dataset for the analysis of functional brain network connectivity, represents vision.
In this study, VBN is constructed using both MC and PC as FC feature which is derived from extracted TS of each active voxel. Each TS is detrended and z-score-normalized before it is used for subsequent analyses. Augmented Dickey Fuller test is executed to check the stationarity of TS. All statistical tests which are reported in this paper, conducted at 5\% significance level.

\subsection{Comparison between $FC^{+}$ and $FC^{-}$ based on MC and PC}
\label{sec:result1}

\begin{figure}[t!]
\centering
\includegraphics[width=3.5in]{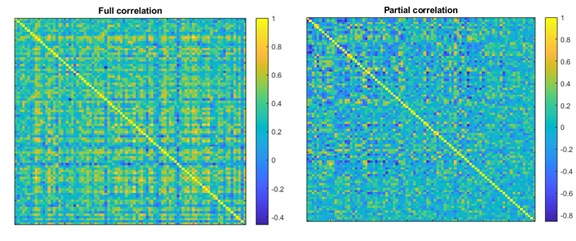}
%\DeclareGraphicsExtensions 
\caption{Marginal (\textit{left}) and Partial correlation (\textit{right}) as obtained for one representative subject. A lot of positive connectivity was seen to be either removed or lessened in strength in PC, as expected, since PC captures the direct linear association between network nodes after eliminating the spurious effects from all confounders.}
\label{fig_figure4}
\end{figure}

VBN is formed based on MC and PC between TS, taken pairwise, exhaustively across all active voxels. For each run, MC and PC matrices are computed. Due to the short duration of the BOLD hemodynamic response, in this study, MC is computed only at zero-lag \cite{ref2}. 
The visualization of MC and PC as obtained for one session of a representative subject is shown in Figure~\ref{fig_figure4}. 
Since many researchers are often interested to investigate the significant connections in brain network, in stead of analyzing the entire correlation that includes both significant and insignificant edges, our study considers only significant connections for further analysis. In this study, the \textit{significant connectivity} is defined as, \textit{"selected network edges which are comparatively stronger (above a pre-defined threshold) and consistent in both MC and PC"}. Fisher Z transformation is computed on both MC and PC values and then significant network edges are extracted if the values in both correlation measures are either greater than 0.5 or lesser than 0. In our study, since we were interested to construct VBN with significantly strong positive connectivity, threshold value was set to 0.5 in order to eliminate all weaker interaction between network nodes.   
The edges which are significantly consistent in both MC and PC will yield two types of brain network, as described in Section~\ref{sec:Construction of VBN and Connectivity Matrix}: positive threshold is used to construct positive visual connectivity network or $FC^{+}$ while the negative threshold is used to construct negative visual connectivity network or $FC^{-}$, for each session. A total of 100 such positive and negative VBNs are formed across all subjects, which are then used for graph-theoretical analysis. $FC^{+}$ and $FC^{-}$ as obtained for one session of a representative subject is shown in Figure~\ref{fig_figure5} and the corresponding positive and negative connectivity matrix is shown in Figure~\ref{fig_figure6}. 

It is seen that, the brain networks with strongest positive connections are denser as compared to brain network with strongest negative connections (see Figure~\ref{fig_figure5}). The comparative statistics regarding total node activation for each subject in $FC^{+}$ and $FC^{-}$ is tabulated in Table~\ref{table_table2}.

\begin{table}[!t]
\caption{Comparative statistics on total number of nodes present $FC^{+}$ and $FC^{-}$ for each subject. N+: total nodes in $FC^{+}$, N-: total nodes in $FC^{-}$}
\label{table_table2}
\centering
\begin{tabular}{|c|c|c|c|c|c|}
\hline
 & \textbf{N+} & \textbf{N-} & \textbf{N+ (\%)} &\textbf{N- (\%)} & \textbf{N+ : N-} \\
 \hline
HC1 &1237 &392 &75.9 &24 &3.16 : 1\\
\hline
HC2 &15212 &3544 &81.1 &18.9 &4.3 : 1\\
\hline
HC3 &862 &178 &82.9 &17.1 &4.84 : 1\\
\hline
HC4 &26843 &3086 &89.7 &10.3 &8.7 : 1\\
\hline
\end{tabular}
\end{table}

\begin{figure}[t!]
\centering
\includegraphics[width=3.6in]{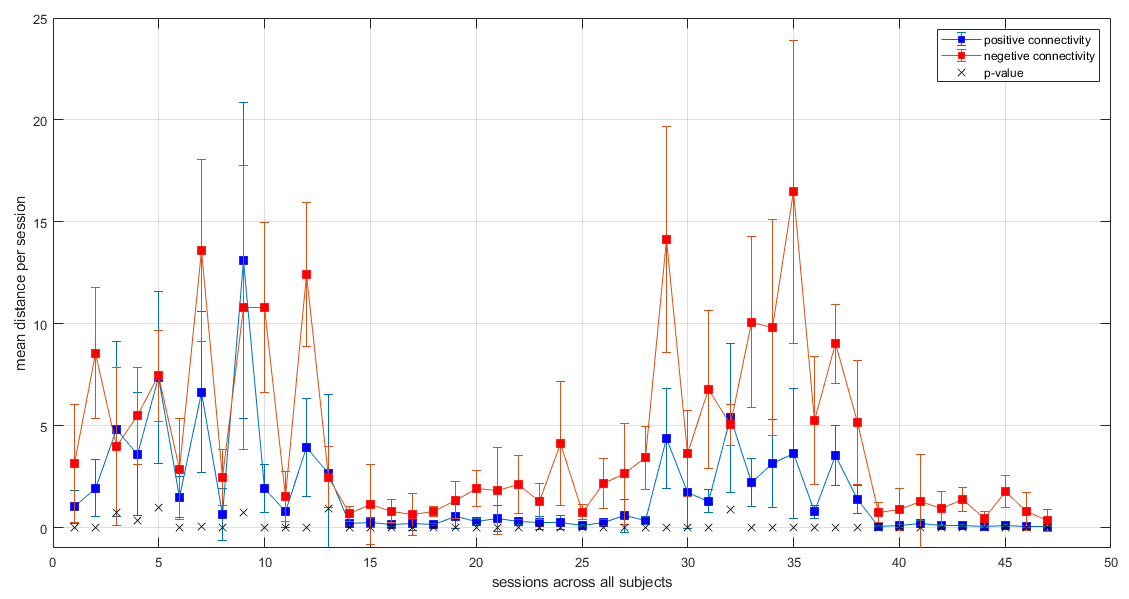}
%\DeclareGraphicsExtensions 
\caption{Average distance between all nodes is plotted for positive brain network (\textit{blue}) and for negative brain network (\textit{red}) across all sessions for all subjects. The p-value (denoted by \textit{black 'X'-mark}) indicates if the difference in mean spatial distance between positive and negative brain networks is statistically significant. As seen from the figure, except 6 sessions (\textit{black 'X'-mark} above zero-line), p-value showed significant difference in spatial distance between positive and negative brain networks, where negatively correlated brain network have more spatial distance between nodes ((\textit{red}) as compared to the positively correlated network (\textit{blue}).}
\label{fig_figure7}
\end{figure}

It is also seen that the strongest negative connections have larger spatial distance as compared to strongest positive connections. This finding is consistent with the previous result as reported in literature \cite{ref26} that showed significant correlation in percentage of negative FC and corresponding spatial distance. In our study, larger distances that occur in $FC^{+}$ are mainly observed in homologous brain locations in the left and right hemisphere. No negative correlation between homologous brain locations is observed in $FC^{-}$.
The average distance profile between each pair of network nodes for $FC^{+}$ and $FC^{-}$ is shown in Figure~\ref{fig_figure7} which clearly indicates longer spatial distance in case of $FC^{-}$. Student paired t-test is conducted on mean distance between $FC^{+}$ and $FC^{-}$, reveals that the difference in mean distance between  $FC^{+}$ and $FC^{-}$ is statistically significant in majority of sessions (42/48) across all subjects. 

\subsection{FC analysis using graph-based features}
\label{sec:FC analysis using graph-based features}

\begin{figure}[t!]
\centering
\includegraphics[width=3.5in]{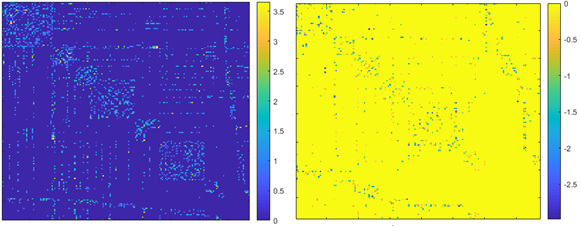}
%\DeclareGraphicsExtensions 
\caption{Weighted adjacency matrix as obtained from $FC^{+}$ (\textit{left column}) and $FC^{-}$ (\textit{right column}) for one session of one representative subject is shown here. As seen from figure, fewer negative connections are found as compared to positive connections which results denser connectivity in case of positively correlated brain network than negatively correlated brain network.}
\label{fig_figure6}
\end{figure}

Statistical tests are also on extracted features in order to check the consistency in statistically significant differences between $FC^{+}$ and $FC^{-}$ across all subjects. Brain-functional connctivity toolbox (http://www.brain-connectivity-toolbox.net) is utilized in order to extract various graphical features \cite{ref22}. Undirected weighted adjacency matrix as obtained from network graph for one session of a representative subject is shown in Figure~\ref{fig_figure6}.

\subsubsection{On whole time series:}
\label{sec:wholeTS}
\begin{figure}[t!]
\centering
\includegraphics[width=3.3in]{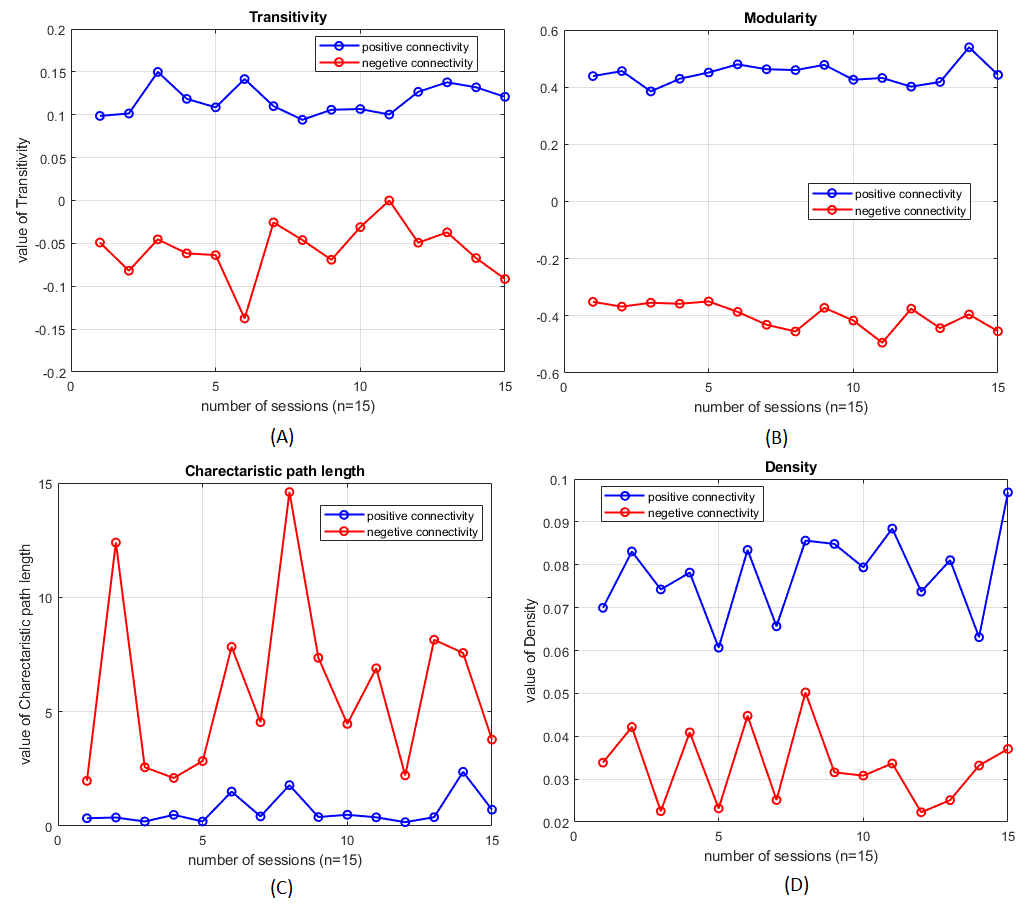}
%\DeclareGraphicsExtensions 
\caption{Global graph-based measures for one representative subject- showing how the global properties varies between positively correlated (\textit{blue}) and negatively correlated (\textit{blue}) VBN. Four global properties of network are computed; Transitivity (\textbf{A}), Modularity (\textbf{B}) quantifies network segregation, whereas Characteristic path length (\textbf{C}) and Density (\textbf{D}) quantifies network integration. Higher Density with shorter Characteristic path length was observed in positively correlated network. On the other hand, negatively correlated network is characterized by lower Density and higher Characteristic path length.}
\label{fig_figure8}
\end{figure}

Four global features, Transitivity (T), Modularity (Q), Characteristic path length ($\lambda$) and Density (D) as described in Section~\ref{sec:globalfeatures} are extracted from $FC^{+}$ and $FC^{-}$ for each session. Figure~\ref{fig_figure8} shows the differences in global characteristics between $FC^{+}$ and $FC^{-}$ for one representative subject. In $FC^{+}$, Analysis of variance (ANOVA) showed significant differences for all features across all subjects. This results due to high inter-subject variability with respect to total number of voxel activation at each fMRI sessions. However, in $FC^{-}$, the difference in T and $\lambda$ for all sessions are found to be statistically insignificant across all subjects. This reflects the significance of anti-correlation in analyzing VBN by T and $\lambda$. While comparing global features between $FC^{+}$ and $FC^{-}$, the difference in all global features T, Q, $\lambda$ and D  are found to be statistically significant for all four subjects. \par

\begin{figure}[t!]
\centering
\includegraphics[width=3.5in]{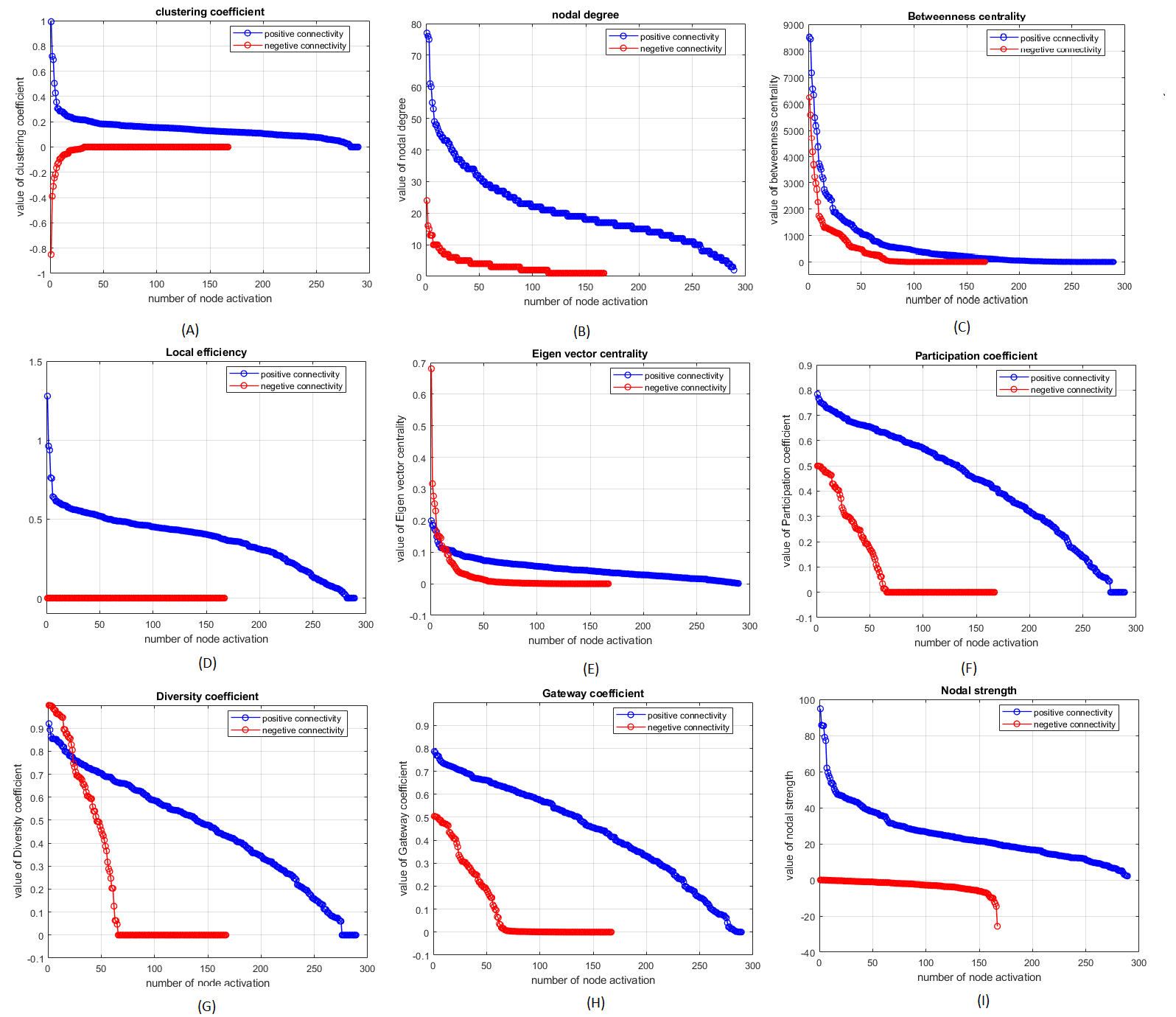}
\caption{The distribution of each local graph-based feature of one particular session is plotted for one representative subject. The figure illustrates how local properties computed at each node of network varies between positively correlated (\textit{blue}) and negatively correlated (\textit{red}) VBN. Among all extracted local features, for all sessions, statistically significant differences between positive and negative network were found for Clustering coefficient, Degree, Participation coefficient and Gateway coefficient and Strength and this is seen to be consistent across subjects. The plot is obtained by sorting the local feature values in descending order for both $FC^{+}$ and $FC^{-}$.}
\label{fig_figure9}
\end{figure}

Nine local features as described in Section~\ref{sec: localfeatures} are extracted from each network graph. The distribution of local graph features of a session for one representative subject is plotted in Figure~\ref{fig_figure9}, showing the high variation between $FC^{+}$ (blue) and $FC^{-}$ (red). Student-paired t-test on local feature 'Strength' showed significant differences between $FC^{+}$ and $FC^{-}$ across all four subjects among all local features. Apart from Strength, three out of four subjects have shown consistently significant differences in CC, Degree, PCf and GC between $FC^{+}$ and $FC^{-}$, across all sessions.

\subsubsection{On image complexity-specific time series}
\label{sec:On image complexity-specific time series}

\begin{figure}[t!]
\centering
\includegraphics[width=3.5in]{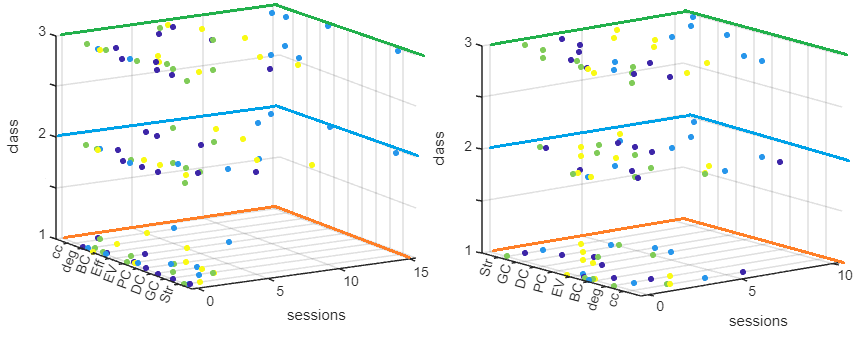}
%\DeclareGraphicsExtensions 
\caption{Comparison of graph theoretical local features that yield significant statistical differences between COCO vs. ImageNet (class-1), ImageNet vs. SUN (class-2) and COCO vs. SUN (class-3) for $FC^{+}$ (\textit{Left}) and $FC^{-}$ (\textit{Right}). As seen from the figure, across all subjects and all sessions, the number of features that yield significant differences in local graphical properties is more in case of class-2 and class-3, as compared to class-1 in both $FC^{+}$ and $FC^{-}$. This captures how brain network architecture changes when images from different complexities are viewed.}
\label{fig_imagewise1}
\end{figure}

The considered global and local features are also extracted from the three image complexity-specific TS that represent COCO, ImageNet and SUN (Figure~\ref{fig_figure2}), in order to examine the differences in properties of VBN, constructed from these three distinct TS. Student paired t-test is conducted on each graphical features in order to check if any changes in graphical properties exist between VBN of (i) COCO vs. ImageNet (class-1), (ii) ImageNet vs. SUN (class-2) and (iii) COCO vs. SUN (class-3), which is statistically significant. No global features across sessions have shown statistical significant differences between the three classes. This is due to the high variability in number of voxel activation across sessions of a specific subject, that yields high differences in global network characteristics across all sessions. Contrary to this, some local properties of these three brain networks (i.e. COCO, ImageNet and SUN) have shown statistical significant differences at each session. The total number of local features that showed significant differences across all sessions and all subjects are visualized by swarm chart in Figure~\ref{fig_imagewise1}. As seen from this figure, across all sessions and across all subjects, there are more number of features that showed significant differences in local graphical properties of VBN of ImageNet vs. SUN (class-2) and COCO vs. SUN (class-3). However, there are very few local features whose difference are seen to be statistically significant between ImageNet and COCO (class-1).

\paragraph{\textbf{Classification of VBNs: ImageNet vs. COCO vs. SUN}}

In order to validate the result of this statistical findings, as described in Section~\ref{sec:On image complexity-specific time series}, XGBoost classification is performed, aiming to differentiate the brain networks of ImageNet, COCO and SUN using extracted local graph features. XGBoost (Extreme Gradient Boosting) is a powerful ensemble machine learning algorithm, that combines the predictions of multiple base learners to produce a stronger prediction \cite{ref47}. XGBoost uses distributed, scalable gradient-boosted decision trees that provides parallel tree boosting and instead of providing a single decision at each leaf node of the decision tree, it includes real-value scores of whether an instance belongs to a specific class. The decision is made by converting the scores into categories using a certain threshold only after the tree reaches to its maximum depth. In this study, the nine local features as obtained from VBN are used for XGBoost classification. The performance of the model is evaluated by splitting the data, 80\% for training, 10\% for validation and the rest 10\% for testing. No data leakage
is allowed between three splits. With k-fold (k=10) cross validation, in case of $FC^{+}$, the average accuracy of 86.5\% is obtained for 3-class classification across all four subjects. In binary classification, the average accuracy across subjects are  88.5\%, 91.5\% and 90.75\% in classifying (i) ImageNet vs. COCO, (ii) ImageNet vs. SUN, and (iii) COCO vs. SUN, respectively. In case of $FC^{-}$, XGBoost classification yields average accuracy of 84.25\% (3-class), 85.5\% (ImageNet vs. COCO), 89.5\% (ImageNet vs. SUN) and 89\% (COCO vs. SUN). The performance measures of XGBoost classification are tabulated in Table~\ref{table_table6}.
The performance of visual network classification across visual datasets using the present graph theoretical approach is comparable with our previous studies \cite{refourtopology,refourtencon} that have used network topological features as computed from persistent homology \cite{refourtopology} and deep neural network architectures \cite{refourtencon} for classifying these visual networks across the considered visual datasets.

\subsection{Unveiling central nodes in VBN}
\label{sec:centralnode}

Researchers are often interested to identify the most important or relevant node in a network. This key node is referred as the central node of a network. In the context of brain functional connectivity network, a central node is defined as \textit{the most connected one through which more information passes within a brain network} \cite{ref1}. Thus, in the current study from each local features, the node corresponds to highest nodal value is identified for all sessions as the central node, which is considered to be most significant. This is done by first sorting the nodes according to their feature score, and then, the first ranked node is extracted, which is considered as the most relevant one with respect to the specific graph feature. Being most influential, it is inferred that the identified  central node in $FC^{+}$ and $FC^{-}$ is associated with the highest functional connectivity as it participates in maximum information transfer. The central nodes corresponds to the highest nodal values are identified from each graph-based local measures for all the sessions in both $FC^{+}$ and $FC^{-}$. The detection of central nodes will help in identifying the most important visual processing region in the brain network. Moreover, the nature of correlation exists between local features of central nodes can also be established. These are discussed in subsequent subsections.

\subsubsection{Relationship between local features of central nodes}

Here, we study the correlation among different features corresponding to central node values across all sessions for each subject. For nine local features, a total of 36 different correlations are obtained for each subject, out of which several significant strong correlations are observed for all four subjects. To statistically prove this observation, a linear model is fitted for each type of significant local features. The strong correlation among features which are seen to be consistent across all four subjects are identified and tabulated in Table~\ref{table_table4}.

\paragraph{\textbf{Relationship consistency in $FC^{+}$ across subjects}}

As seen from Table~\ref{table_table4}, among all local features, taken pairwise, we have found 4 feature pairs that have shown statistically significant strong correlations for $FC^{+}$, which are consistent across all 4 subjects. These are- (i) BC and nodal Degree ($mean R^{2}(adj) =0.77$), (ii) EVC and nodal Degree ($mean R^{2}(adj)=0.645$), (iii) EVC and BC ($mean R^{2}(adj)=0.668$) and (iv) DC and PCf ($mean R^{2}(adj)=0.765$).

\paragraph{\textbf{Relationship consistency in $FC^{-}$ across subjects}}
For $FC^{-}$, we have found 2 feature pairs that have shown statistically significant strong correlations which are consistent across all 4 subjects. These are (i) nodal Strength and nodal Degree ($mean R^{2}(adj)=0.855$)and (ii) DC and PCf ($mean R^{2}(adj)=0.99$). \par

\paragraph{\textbf{Comparison of positively and negatively correlated central node local features across $FC^{+}$ and $FC^{-}$}}

\begin{figure*}[t!]
\centering
\includegraphics[width=6.5in]{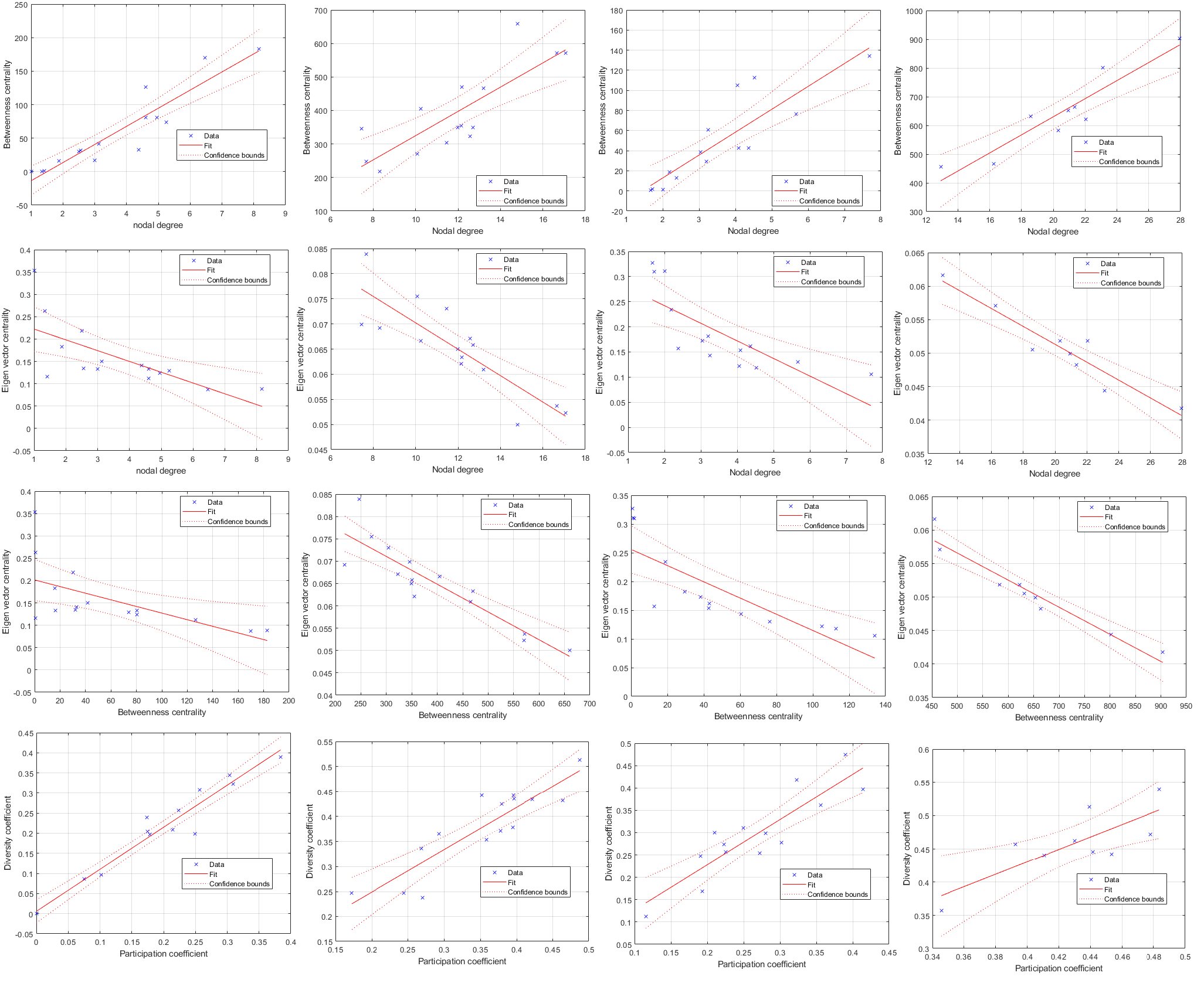}
\caption{Relationship among local features of central node for all sessions that yield consistent result for all subjects are plotted for positively correlated VBN ($FC^{+}$). In the figure, each column represents the relationship for each subject. The strong positive correlation is observed between (i) Betweenness centrality and nodal degree (\textit{Row-1}), (ii)  Diversity coefficient and Participation coefficient (\textit{Row-4}). Strong negative correlation is observed between (i) Eigen vector centrality and nodal Degree (\textit{Row-2}) and (ii) Eigen vector centrality and Betweenness centrality (\textit{Row-3}).}
\label{fig_figure10a}
\end{figure*}

\begin{figure*}[t!]
\centering
\includegraphics[width=6.5in]{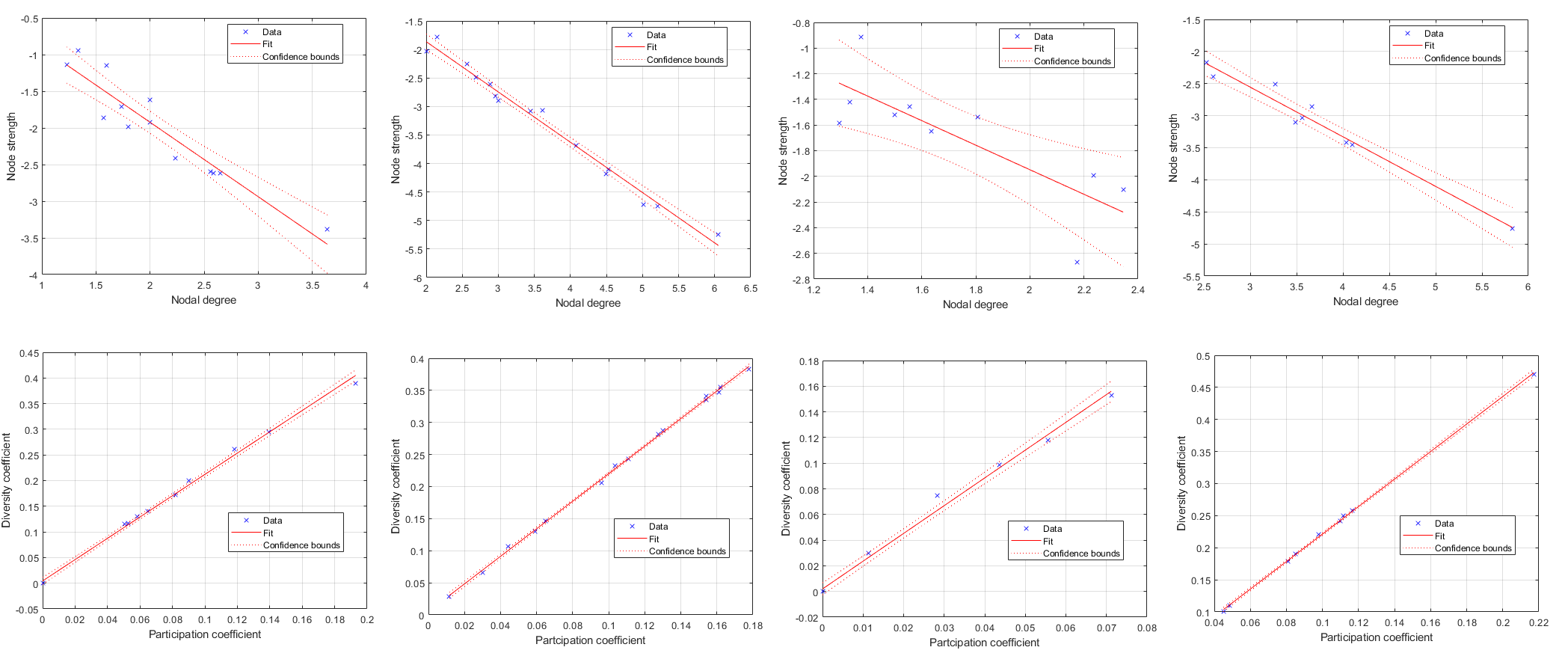}
\caption{Relationship among local features of central node for all sessions that yield consistent result for all subjects are plotted for anti-correlated VBN ($FC^{-}$). Similar to \ref{fig_figure10a}, each column represents the relationship for each subject. The strong negative correlation is observed between nodal Strength and Degree \textit{Row-1}. Similar to positively correlated brain network, strong positive correlation is obtained between Diversity coefficient and Participation coefficient \textit{Row-2}.}
\label{fig_figure10b}
\end{figure*}

On comparing the correlations obtained for $FC^{+}$ and $FC^{-}$, it is seen that the consistently strongest correlations across all subjects were obtained for $FC^{-}$ as compared to $FC^{+}$. For $FC^{-}$, $R^{2}(adj)$ ranges from 0.88 to 0.999 across all subjects (except sub-3 for strength and degree relationship). For $FC^{+}$, $R^{2}(adj)$ values are found to be highly scattered (ranges from 0.33 to 0.95) across all subjects. This leads us to the hypothesis that anti-correlation is important for understanding VBN.
Strongest positive correlations are found for the pair: BC and nodal Degree in $FC^{+}$. On the other hand, the correlation between the feature pair: DC and PCf is seen to be very strong in both $FC^{+}$ ($mean R^{2}(adj)=0.77$) and $FC^{-}$ ($mean R^{2}(adj)=0.99$). 
EVC is found to be negatively correlated with both nodal Degree and BC for $FC^{+}$. Significant strong negative correlation in $FC^{-}$  is obtained between nodal Strength and Degree. The relationship between local features of central nodes are illustrated in Figure~\ref{fig_figure10a} and Figure~\ref{fig_figure10b} for $FC^{+}$ and $FC^{-}$ respectively.

\begin{table}[!t]
\caption{Relationship between local features of central nodes in positive and negative VBN which are seen to be consistent for all subjects. These are Betweenness centrality (BC), Eigen vector centrality (EVC), Diversity coefficient (DC), and Participation coefficient (PCf)}
\label{table_table4}
\centering
\begin{tabular}{|c|c|c|c|c|c|}
\hline
\textbf{VBN} & \textbf{Local} & \textbf{Sub} & \textbf{r-value} &\textbf{p-value} & \textbf{${}R^{2}(adj)$} \\
 &\textbf{feature} & & & &\\
\hline

$FC^{+}$ & BC vs.  & HC1  & 0.9281	&6.11e-07	&0.851 \\
\cline{3-6}
 &Degree  & HC2	& 0.8138	& 0.00022	& 0.636	 \\
\cline{3-6}
  &  & HC3	& 0.8729	& 4.6e-05	& 0.742	 \\
\cline{3-6}	
  &  & HC4	& 0.9336	& 0.00023	& 0.853	 \\
\cline{2-6}
 & EVC Vs.  & HC1  & -0.6931	&0.0041	&0.44 \\
\cline{3-6}
 &Degree & HC2	& -0.8522	& 0.055e-3	& 0.705	 \\
\cline{3-6}
 &  & HC3	& -0.7692	& 0.0013	& 0.558	 \\
\cline{3-6}
 & & HC4	& -0.9451	& 0.00012	& 0.878	 \\
\cline{2-6}
& EVC Vs.  & HC1  & -0.6203	&0.0136	&0.338 \\
\cline{3-6}
 &BC & HC2	& -0.8958	& 6.6e-06	& 0.787	 \\
\cline{3-6}
 & & HC3	& -0.8093	& 0.0004	& 0.626	 \\
\cline{3-6}
 & & HC4	& -0.9662	& 2.25e-05	& 0.924	 \\
\cline{2-6}
& DC Vs.  & HC1  & 0.9771	&4.05e-10	&0.951 \\
\cline{3-6}
 &PCf & HC2	& 0.8855	& 1.13e-05	& 0.768	 \\
\cline{3-6}
 & & HC3	& 0.8871	& 2.34e-05	& 0.769	 \\
\cline{3-6}
 & & HC4	& 0.7926	& 0.0108	& 0.575	 \\
\cline{2-6}
\hline
$FC^{-}$ & Strength Vs.  & HC1  & -0.9439	&1.25e-06	&0.881 \\
\cline{3-6}
 &Degree  & HC2	& -0.9916	& 5.9e-13	& 0.982	 \\
\cline{3-6}
  &  & HC3	& -0.7984	& 0.0056	& 0.6	 \\
\cline{3-6}	
 &  & HC4	& -0.9821	& 2.4e-06	& 0.96	 \\
\cline{2-6}
 & DC Vs.  & HC1  & 0.9984	&3.58e-15	&0.997 \\
\cline{3-6}
 &PCf & HC2	& 0.9994	& 1.08e-20	& 0.999	 \\
\cline{3-6}
 & & HC3	& 0.9968	& 4.78e-10	& 0.993	 \\
\cline{3-6}
 & & HC4	& 0.9997	& 1.03e-12	& 0.999	 \\
\cline{2-6}
\hline
\end{tabular}
\end{table}

%\begin{figure*}[t!]
%\centering
%\includegraphics[width=6.5in]{images/figure10new.png}
%\caption{Relationship among local features of central node for all sessions that yield consistent result for all subjects are plotted. In the figure, each column represents the relationship of each subject. 
%For positively correlated network (\textit{blue}), strong positive correlation was observed between (A) Betweenness centrality and nodal degree and (D) Diversity coefficient and Participation coefficient. Strong negative correlation was observed between (B) Eigen vector centrality and nodal Degree and (C) Eigen vector centrality and betweenness centrality. For negatively correlated network (\textit{red}), strong negative correlation was obtained between (E) nodal Strength and Degree. Similar to positive brain network, strong positive correlation was obtained between (F) Diversity coefficient and Participation coefficient.}
%\label{fig_figure10}
%\end{figure*}

\subsubsection{Spatially scattered central node locations in $FC^{-}$}
\label{sec:central_node_locations}
\begin{figure}[t!]
\centering
\includegraphics[width=3.3in]{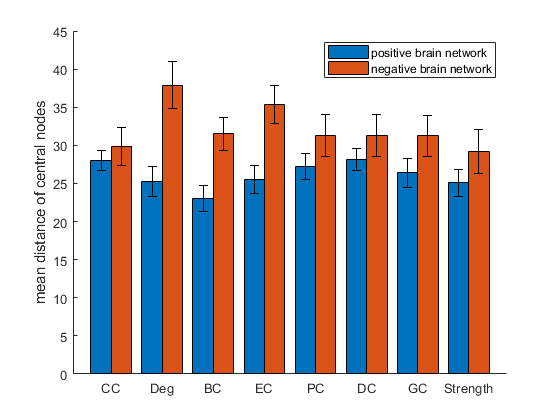}
%\DeclareGraphicsExtensions 
\caption{The mean distance among central nodes for all sessions are shown for positively correlated network (\textit{blue}) and negatively correlated network (\textit{red}) reveals that central nodes are scattered in negative brain network as compared to brain network with positive connectivity. This directly reflects the observation that we have seen in Figure~\ref{fig_figure7}- demonstrating higher spatial distance in case of negative brain network ($FC^{-}$) as compared to positive brain network ($FC^{+}$).}
\label{fig_figure11}
\end{figure}

\begin{figure}[t!]
\centering
\includegraphics[width=3.3in]{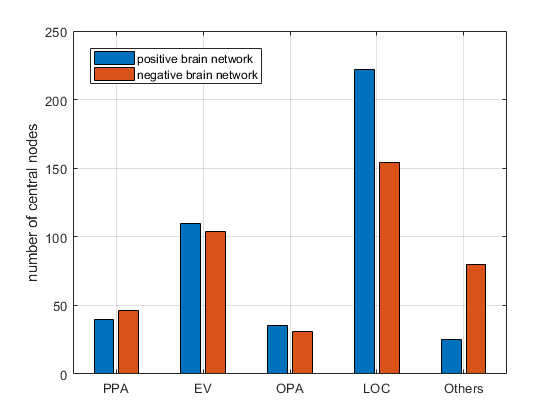}
%\DeclareGraphicsExtensions 
\caption{Number of locations of central nodes for all subjects obtained in sub-cortical brain regions that represents vision. The central nodes are seen to be distributed among four visual ROIs: PPA, EV, OPA and LOC. The 'others' in the figure represents regions outside the brain visual ROIs, considered in this study. Comparing all visual ROIs, maximum number of central nodes are found at the object-selective region LOC.}
\label{fig_figure12}
\end{figure}

As discussed in Section~\ref{sec:centralnode} central node of a network helps to identify the most important ROIs in the brain that represents vision. The 5 visual processing ROIs as defined in \cite{ref14} are selected for the current study. These are scene selective ROIs: PPA, RSC, OPA and object selective ROIs: LOC and EV (the classical V1 or V2 region). 
For each local feature, the central node locations are identified for all sessions and for all subjects. This is shown in Figure~\ref{fig_figure12}. It is seen that the locations of central node obtained for all features in all sessions are highly scattered across the 5 visual processing ROIs in case of $FC^{-}$ as compared to $FC^{+}$ for all subjects. On the contrary, in case of $FC^{+}$ the locations of central node are seen to be highly localized in specific visual processing ROIs across all subjects. This visual observation is further quantified by measuring the mean Euclidean distance among network nodes. This is shown in Figure~\ref{fig_figure11}. The figure clearly depicts that the mean distance among central nodes are larger and hence are more likely to be scattered in $FC^{-}$ than in $FC^{+}$. \par

The number of central nodes found at each visual ROI for each feature is tabulated in Table~\ref{table_table5}. Across all features the maximum number of central nodes are found at LOC in both $FC^{+}$ and $FC^{-}$. The prevalent understanding that LOC plays an important role in human object recognition is further strengthened from this result. Followed by LOC, the second highest central nodes are found at EV. This is illustrated in Figure~\ref{fig_figure12}. 
Comparing $FC^{+}$ and $FC^{-}$, some central nodes are found in regions other than the considered visual processing areas as described in Section~\ref{sec:central_node_locations}. These nodes are referred as "others" in our study. However, it is seen that the total number of these "others" central nodes are relatively lesser in case of $FC^{+}$ as compared to $FC^{-}$. 
This is possibly due to the presence of very few positive correlations that are observed between the considered visual ROIs and other brain regions in $FC^{+}$. However, in case of $FC^{-}$, more negative correlations are obtained between these visual ROIs and "other" brain regions. As a result, more number central nodes in "others" brain region are found in $FC^{-}$ as compared to $FC^{+}$ (Figure~\ref{fig_figure12}).

\begin{table*}[!t]
\caption{Total number of locations of central nodes as obtained for each local features for all subjects is tabulated. As seen from the table, the maximum number of central nodes are found in LOC, followed by EV.}
\label{table_table5}
\centering
\begin{tabular}{|c|c|c|c|c|c|c|c|c|c|c|}
\hline
 &\multicolumn{2}{c|}{\textbf{PPA}}  &\multicolumn{2}{c|}{\textbf{EV}} & \multicolumn{2}{c|}{\textbf{OPA}} &\multicolumn{2}{c|}{\textbf{LOC}} & \multicolumn{2}{c|}{\textbf{Others}} \\
 
\hline
 & \textbf{$FC^{+}$} & \textbf{$FC^{-}$} & \textbf{$FC^{+}$} & \textbf{$FC^{-}$} & \textbf{$FC^{+}$} & \textbf{$FC^{-}$} & \textbf{$FC^{+}$} & \textbf{$FC^{-}$} & \textbf{$FC^{+}$} & \textbf{$FC^{-}$}\\
 \hline
CC &7 &4 &11 &16 &4 &8 &26 &15 &6 &9\\
\hline
Degree &4 &3 &14 &13 &4 &0 &30 &16 &2 &20\\
\hline
BC &5 &5 &20 &14 &3 &1 &25 &23 &1 &9\\
\hline
EVC &4 &7 &13 &10 &6 &1 &27 &21 &4 &13\\
\hline
PC &5 &6 &10 &12 &4 &8 &32 &19 &3 &7\\
\hline
DC &5 &6 &12 &12 &4 &8 &29 &19 &4 &7\\
\hline
GC &5 &10 &14 &10 &5 &4 &28 &18 &2 &9\\
\hline
Strength &5 &5 &16 &17 &5 &1 &25 &23 &3 &6\\
\hline
 \textit{Total} &40 &46 &110 &104 &35 &31 &\textbf{222} &\textbf{154} &25 &80\\
\hline
\end{tabular}
\end{table*}

\begin{table*}[!t]
\caption{Results of image complexity-specific VBNs classification using local graphical features. The performance measures of XGBoost classifier across all fMRI sessions are reported for all subjects. The prediction result reflects how VBN properties differ while viewing different images of different complexities. As seen from the table, the consistently high accuracy is obtained while distinguishing SUN from ImageNet and COCO across all four subjects.}
\label{table_table6}
\centering
\begin{tabular}{|c|c|c|c|c||c|c|c|c|}
\hline
  &\multicolumn{4}{c|}{\textbf{Positive network ($FC^{+}$)}}  &\multicolumn{4}{c|}{\textbf{Negative network ($FC^{-}$)}} \\
 \cline{2-9}
  &\textbf{Precision} & \textbf{Recall} & \textbf{F1 score} & \textbf{Accuracy\%} & \textbf{Precision} & \textbf{Recall} & \textbf{F1 score} & \textbf{Accuracy\%}\\
 \hline
3-class &86.5	&86	&86.5	&86.5		&84.75	&84	&84.25	&84.25 \\
\hline
ImageNet vs. COCO &88	&88.25	&89.5	&88.5		&85.75	&85.25	&85.25	&85.5\\
\hline
ImageNet vs. SUN &91.5	&90.25	&91	&\textbf{91.5}		&88.5	&88.5	&88	&\textbf{89.5}	\\
\hline
COCO vs. SUN &90.5	&90.75	&90.5	&\textbf{90.75}	   &88.75	&88.5	&88.5	&\textbf{89}	\\
\hline

\end{tabular}
\end{table*}

\section{Discussion}
Vision science, particularly machine vision, has been revolutionized by introducing large-scale image datasets and statistical learning approaches. However, human neuroimaging studies of visual perception is still remain the fundamental open problems. There are very limited literature available that studied the VBN on a large-scale image dataset. The release of a publicly available dataset “BOLD5000” \cite{ref14}, that contains fMRI scans of subjects acquired while viewing over 5000 images, has made it possible to study the brain dynamics during visual tasks in greater detail. This motivates us to explore the FC pattern of brain network that represents vision. In this regard, the graph theoretical analysis have established a mathematical framework that reveals meaningful information about the graphical properties of human brain networks by modelling pairwise communications between network nodes. The main focus of this study is to investigate the nature and relationships of brain functional properties that emerge through interactions of neurons during visual task using fMRI TS information. \par
The activities in brain regions obtained from fMRI reveals an organized structure with segregation and integration properties \cite{ref1}. The neuronal activity in brain can be measured in two ways; (i) mutual activation which corresponds to positive correlations, when two brain regions activate at the same time, and (ii) dissociated activation, which corresponds to negative correlations or anti-correlations that occurs when one region is more active than the other region which is less active. While most of the studies on brain network analysis utilized positive correlations of network elements, very less attention has been paid to brain network that involves negative edges. This is because negative FC in the context of network physiology are less understood and have been a subject of debate since several studies demonstrated that the negative correlation could be a result of artifact introduced by a global signal regression procedure during pre-processing of fMRI data \cite{ref31,ref32,ref33}, thereby needs to be meticulously interpreted. However, some recent literature have reported the significance of negative correlations that strongly suggest that the presence of negative connectivity should not be simply discarded since they may contain neurobiologically relevant information \cite{ref27,ref28,ref29,ref30}. Thus, the first contribution of the present study is the inclusion of anti-correlation to examine the properties of negative VBN, along with the exploration of positive connectivity. Neurobiological inferences are drawn by utilizing graph theoritical measures on both of these positive and negative functional brain networks, that provides an insight in understanding VBN dynamics.\par
The second contribution of this work is the use of PC in constructing VBN in order to capture direct (true) connectivity between network nodes (see Section~\ref{parcorr}). In spite of having great potential in studying direct FC, the application of PC in the neuroimaging community has been limited since the estimation of PC is little challenging as compared to MC. Direct estimation of PC based on the regression approach is ineffective with respect to computational time and may fail in presence of multicollinearity among TS of network nodes. In such cases, PCs can be estimated by computing precision matrix (the inverse of the covariance matrix) \cite{ref34}. However, estimation of the precision matrix is not trivial in case of large dimension, as the computation includes inverse of covariance matrix. Especially, it fails when the number of observations are less than the number of nodes in a network \cite{ref36} which is very common particularly in fMRI studies. Moreover, precision matrix in PC needs to satisfy the positive definite criterion which increases complexity in its estimation. To overcome this difficulty, Moore-Penrose pseudo inverse of the covariance matrix \cite{ref42}- which is one of the most popular approach in estimating PC is used in this work. Using this PC, VBN is formed by considering consistently and significantly strong network edges (see Section~\ref{sec:result1}). It is observed that significant MCs were eliminated in PC matrix that occurred due to confounding factors  in MC (Figure~\ref{fig_figure4}). Similarly, several negative connections  disappeared in PC and what remain constitute the $FC^{-}$ tend to have a strong neurobiological significance. \par
The third contribution of this study is to execute graph-theoretical analysis on PC based different kinds of VBN (see Section~\ref{sec:FC analysis using graph-based features}). The first analysis of this contribution reveals significant differences in network properties between $FC^{+}$ and $FC^{-}$. In the second analysis, the graph-based network properties are studied for VBN, constructed from image complexity-specific TS representing COCO, ImageNet and SUN for both $FC^{+}$ and $FC^{-}$ (see Section~\ref{sec:On image complexity-specific time series}. In order to differentiate these three distinct brain networks, local graph-based features are utilized. XGBoost, a currently popular ML classifier, which has been already showed its high efficacy in prediction, is used for this purpose.  The result of this classification for $FC^{+}$ and $FC^{-}$ is shown in Table~\ref{table_table6}. Across all subjects, improved accuracy is obtained while classifying brain network of SUN from that of ImageNet and COCO. The classification accuracy of brain networks of COCO and ImageNet is found to be relatively lesser, may be due to presence of common objects in both of these datasets. One example of such spatial context similarity is shown in Figure~\ref{fig_imgs}; the baseball ground image in COCO and ImageNet that contain similar information, leading to high probability of misclassification.  On the other hand, the images which were shown to the participants from SUN database were more scenic with less focus on any particular object, which leads to better results when brain networks of SUN and brain networks of ImageNet and COCO are classified. Similar observation is seen while classifying ImageNet, COCO and SUN  using negative correlated brain networks $FC^{-}$. However,the overall classification accuracy of $FC^{-}$ is found to be decreased by 2\% as compared to that of $FC^{+}$. \par
Apart from studying the differences in graphical properties of different types of VBNs, the associations of these properties with each other based on central node across sessions, are also investigated (see Section~\ref{sec:centralnode}). One interesting observation while comparing this relationship for $FC^{+}$ and $FC^{-}$ is that, consistently stronger correlation is obtained in case of $FC^{-}$ ($R^{2}$ varires in the range 0.88-0.99) for all subjects as compared to $FC^{+}$ ($R^{2}$ varies in the range 0.33-0.95). All these findings again highlight the importance of considering anti-correlation in understanding VBN. \par
While comparing the five visual ROIs, maximum number of central nodes are found in object selective region (brain regions that are activated while viewing objects) LOC, followed by EV (Figure~\ref{fig_figure12}). Compared to LOC and EV, a small fraction of central nodes are found in scene selective regions (brain regions that are activated while viewing scenes) like OPA and PPA. This may be due to the reason that majority of the images that the participants viewed contain objects, either singular or multiple, rather than scene images. In BOLD5000 database, the object-based images and scene-based images which were viewed by each participant during each run of the experiment was in the ratio of 3.916:1.  This result is consistent in both $FC^{+}$ and $FC^{-}$ across all subjects; thereby signifying the importance of considering anti-correlated network in understanding how vision represents in brain. However, generalization of all these findings require more number of participants, sessions and diversity of stimulated images, that will help advance understanding of human visual functional neural networks. Although 5,000 images are quite large to study brain visual dynamics using fMRI, it is still relatively smaller when compared to human visual experience in everyday life.

\section{Conclusion}
The paper outlines a comprehensive graph-theory based brain function connectivity analysis that is performed in order to study the VBN. Though, the variation in the results, obtained for some graph-based features across sessions and subjects is one limitation, that requires further investigation; nevertheless, as a baseline work in understanding VBN, the results showed a good foundation for future fMRI studies on how vision is represented in brain.

\section*{Acknowledgements}
We would like to thank the Mphasis Foundation for the Cognitive Computing grant that supported this research at IIITB.

.%\end{IEEEbiography}

%\begin{IEEEbiography}[{\includegraphics[width=1in,height=1.25in,clip,keepaspectratio]{images/NS.png}}]{Dr. Neelam Sinha}
%is Associate Professor at International Institute of Information Technology (IIIT), Bangalore. She received her Ph.D from Indian Institute of Science (IISc), Bangalore. Her previous stints include MILE Lab, IISc and MR Imaging group at GE Healthcare, Bangalore. Her research interests are in Medical Image Analysis, Computer Vision Strategies for Rapid MRI, Diffusion Weighted MRI, EEG Analysis, Time Series Analysis.
%\end{IEEEbiography}


%\newpage


%\vfill


\end{document}